\documentclass[a4paper]{jpconf}
\usepackage{graphicx}
\begin{document}

\title{Evidence for Intra-Unit Cell Magnetism in Superconducting Cuprates: a Technical Assessment}

\author{Philippe Bourges, Dalila Bounoua, Jaehong Jeong}
\ead{philippe.bourges@cea.fr} 
\address{Laboratoire L{\'e}on Brillouin, CEA-CNRS, Universit\'e Paris-Saclay, CEA Saclay, 91191 Gif-sur-Yvette, France}

% \author{Dalila Bounoua}
% \address{Laboratoire L{\'e}on Brillouin, CEA-CNRS, Universit\'e Paris-Saclay, CEA Saclay, 91191 Gif-sur-Yvette, France}

% \author{Jaehong Jeong}
% \address{Laboratoire L{\'e}on Brillouin, CEA-CNRS, Universit\'e Paris-Saclay, CEA Saclay, 91191 Gif-sur-Yvette, France}

\author{Lucile Mangin-Thro}
\address{Institut Laue-Langevin, 71 avenue des martyrs, 38000 Grenoble, France}

\author{Yvan Sidis}
\address{Laboratoire L{\'e}on Brillouin, CEA-CNRS, Universit\'e Paris-Saclay, CEA Saclay, 91191 Gif-sur-Yvette, France}

\begin{abstract}

Intra unit cell (IUC)  magnetic order observed by polarized neutron diffraction (PND)  is one of the hallmarks of the pseudogap state of high-temperature copper oxide superconductors. This experimental observation, usually interpreted as a result of loop currents, has been recently challenged based on lower statistics data. We here address the crucial issue of polarization inhomogeneities in the neutron beams showing that the original data had a much better reproducibilty. Within these technical limitations, we here propose a self-consistent analysis that potentially solves the controversy. We show that all the reported PND  experiments in superconducting cuprates are actually compatible with the existence of an IUC magnetism. 
\end{abstract}

\section{Introduction}

There has been a long-standing debate among condensed-matter physicists about the origin of the pseudo-gap state in high-temperature superconducting cuprates\cite{revue}. Polarized neutron diffraction (PND) revealed the existence of a Q=0 ordered magnetic phase, hidden inside the pseudo-gap state of four different cuprates families:   $\rm YBa_2Cu_3O_{6+x}$  (YBCO) \cite{Fauque,Mook,Baledent-YBCO,Lucile15,Lucile17},  $\rm HgBa_2CuO_{4+\delta}$ \cite{Li2008,Li2011,Yang2018},   $\rm La_{2-x}Sr_xCuO_4$ \cite{Baledent-LSCO}  and  $\rm Bi_2Sr_2CaCuO_{8+\delta}$ \cite{DeAlmeida-Didry2012,Lucile14}. Interestingly, the ordering temperature 
remarkably  matches both the pseudo-gap temperature T$^*$ and its characteristic hole doping dependence as deduced from resistivity measurements. In strongly underdoped YBCO samples (at doping around 10\%), the measurements  indicate a magnetic order at long range within the experimental limitations of the neutron resolution which establish a correlation  length perpendicular to the CuO$_2$ plane of at least 75 \AA \cite{Fauque,Mook}. However, it appears at short range at higher doping \cite{Lucile15}, suggesting a possible crossover in doping \cite{Lucile14}.

The observed  magnetism can be described as an intra-unit-cell (IUC) magnetic order\cite{CC-review,CC-review2}, breaking time-reversal symmetry
and is usually interpreted in terms of loop currents (LC) order\cite{simon,Varma2006}.  In more recent approaches, the LC order actually occurs as an emergent phenomenon from electronic instabilities at the origin of the pseudogapped metal, such as pair density wave state \cite{Agterberg}, broken SU(2) symmetry \cite{Morice} or a topological order \cite{Chatterjee}. Interestingly,  the neutron IUC magnetism can be as well described by spontaneous ordering of Dirac multipoles \cite{Lovesey,Fechner}. Clearly, the role of the IUC magnetism is one of the most discussed issues in superconducting cuprates. 

Very consistently, PND reports the IUC magnetic order only at specific Bragg positions where the magnetic signal from loop currents is expected and absent otherwise. Typically, the magnetic signal simulated from LCs phase should be maximum around Bragg peaks $\bf Q$=(1,0,L)$\equiv$(0,1,L)  and vanish at larger $\bf Q$ position like  $\bf Q$=(2,0,L)$\equiv$(0,2,L). Further, no magnetic signal is expected at  $\bf Q$=(0,0,L) as there is no net magnestism per unit cell\cite{Fauque,Li2011,CC-review}. In all cuprates, the magnetic nature of the signal was unambiguously proved by neutron polarization analysis \cite{Fauque,Mook,Baledent-YBCO,Lucile15,Li2008,Li2011,Baledent-LSCO,Lucile14}. Further,  it is striking that the YBCO data are nearly indistinguishable from those observed for  $\rm HgBa_2CuO_{4+\delta}$ \cite{Li2008,Li2011,Yang2018} or $\rm Bi_2Sr_2CaCuO_{8+\delta}$ \cite{DeAlmeida-Didry2012,Lucile14}.  All datasets are systematically scaling with their pseudogap doping and temperature dependencies observed in each system. This is remarkable as sample characteristics (mosaicities, masses, shapes,...) differ considerably among these cuprates. Therefore,  the observation of an IUC magnetism, occuring only within the pseudogap state, cannot be coincidental or parasitic. 

Last year, this comprehensive set of data has been questioned by Croft {\it et al}\cite{Hayden} who claim to find  ``no evidence for such orbital loop currents'' in two (millimetric size) YBCO samples. However, that report contains major experimental shortcomings \cite{comment} which prevent Croft {\it et al}\cite{Hayden} to observe the magnetic signal. The primary reason is due to the very small size of their samples. Indeed, they worked on samples at least 100 times smaller than in previous experiments. The counting time was similar in both experiments.  The neutron intensity at the sample position is about 3 times larger in Croft {\it et al}\cite{Hayden}
compared to our measurements (that difference being mostly due to the different reactors nominal power). Clearly, that does not compensate the huge difference in sample mass between both experiments (the neutron intensity being proportional to the sample mass).  As a usual consequence for neutron scattering experiments, their data had insufficient statistics to be able to observe the IUC signal \cite{comment}. Other experimental shortcomings - like  incorrect data calibration, misleading sample comparison and ignorance of the impact of detwinning -  were also crucial limiting factors in their analysis\cite{comment}. Therefore,  none of the data reported by  Croft {\it et al} challenges the universality of the intra-unit cell magnetism in superconducting cuprates. 

In their recent reply\cite{reply} to our comment\cite{comment},  Croft {\it et al} are now pointing out experimental complications related to polarization inhomogeneities in the neutron beam. This issue, that we actually discussed  previously  in a few occasions \cite{Baledent-YBCO,Li2011,Yang2018,Lucile14,CC-review,comment}, needs indeed to be properly addressed but does not dismiss the observation of the IUC magnetism.  We then here discuss possible causes of polarization inhomogeneities in the neutron beam that underline the poorer data quality of Croft {\it et al}  compared to our measurements. We also point out inconsistencies in their data analysis.  Due to these limitations, we  show that Croft {\it et al}\cite{Hayden} data can be as well consistent with the existence of IUC  magnetism.

\section{Normalized spin-flip intensity}

In PND experiment, one measures independently the polarized neutron cross-sections related to different neutron spin states. As emphasized in many reports \cite{Baledent-YBCO,Li2011,Yang2018,Lucile14,CC-review,Hayden,comment,reply}, the relevant quantity to observe the IUC magnetic order is the normalized spin-flip intensity at a Bragg position where the signal is expected. That quantity essentially corresponds to the ratio of two measured neutron intensities $I_{SF}$ and $I_{NSF}$, which stand for the spin-flip and non-spin-flip intensities, respectively. It can be also written as the inverse flipping ratio $R^{-1}_{\rm meas}$: 

%-------------------------------------------------------------
\begin{equation}
\frac{I_{SF}}{I_{NSF}}= R^{-1}_{\rm meas}= \frac{I_{mag}}{I_{NSF}} + R^{-1}_{0}
\label{Imag}
\end{equation}
%-------------------------------------------------------------

\begin{figure}[t]
\includegraphics[width=6.cm,angle=270]{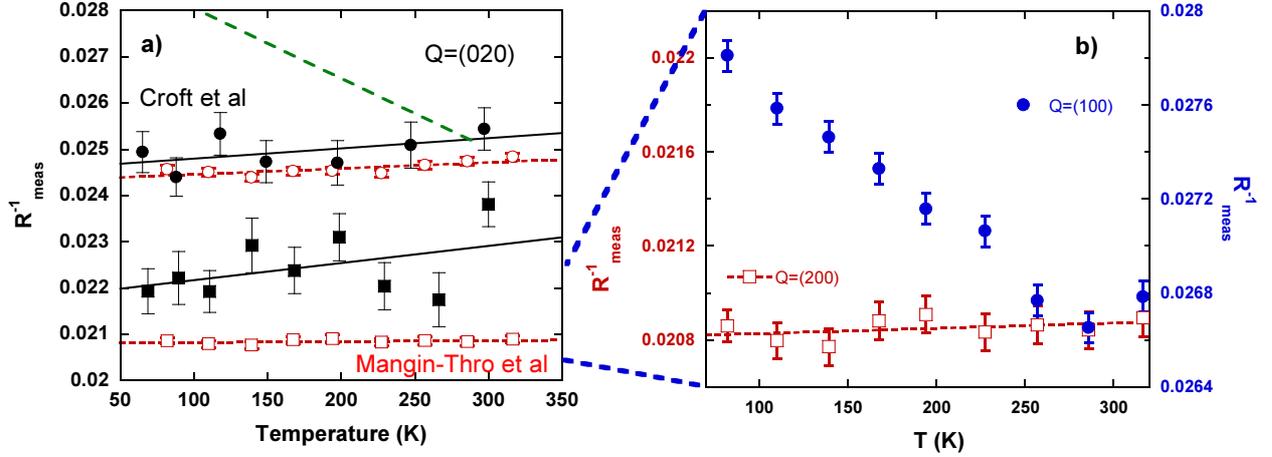}
\caption {
(color online)  a) Bare inverse flipping ratio $R^{-1}_{0}(T)$ versus temperature at Bragg positions  Q=(2,0,0) (open squares) and  Q=(0,2,0) (all other symbols) where the IUC magnetism is absent. This is measured  in three different YBCO samples and two different experimental setups:  YBCO$_{6.6}$ (doping p=0.115) on 4F1/LLB (open symbols) \cite{Lucile17,comment}  (the error bars are lower than the size of the points), YBCO$_{6.54}$ (p=0.104, full circles) and YBCO$_{6.67}$ (p=0.123, full squares) on IN20/ILL \cite{Hayden}.  The  full symbols are the data obtained  ''With re-alignment'' \cite{reply}. The dashed green line represents the data with ''No re-alignment'' \cite{reply}. In all cases, the data points are described by a linear fit. b)  Comparison of $R^{-1}_{0}(T)$  at  Q=(2,0,0) (open squares) and Q=(1,0,0) (full circles) (from \cite{Lucile17}). Note that the data for Q=(200) are the same data as  Fig. \ref{croft-lucile}.a but the vertical scale is zoom in  (as depicted by the dashed blue lines) to scale with the amplitude IUC magnetic signal on Q=(100).   
}
\label{croft-lucile}
\end{figure}

\begin{figure}[t]
\includegraphics[width=5.5 cm,angle=270]{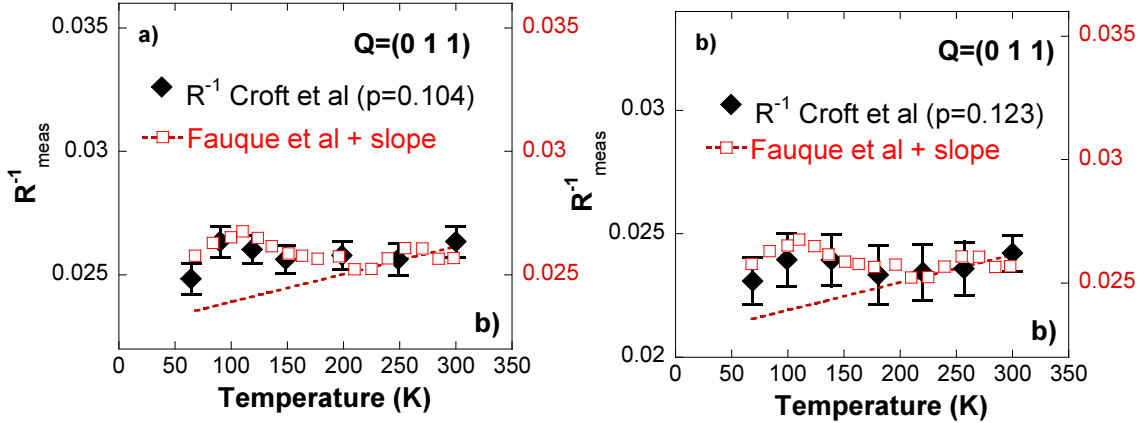}
\caption {
(color online) Inverse of the measured flipping  ratio  $R^{-1}_{0}(T)$ versus temperature at a Bragg position Q=(0,1,1)  where the IUC magnetism is present.  Both figures compare the measurements of two different experiments on 4F1/LLB in YBCO$_{6.6}$ (doping p=0.115) (open symbols)  \cite{Fauque,CC-review} and in two different YBCO samples on IN20/ILL \cite{Hayden}  (full symbols) : a)  YBCO$_{6.54}$ (p=0.104) and b) YBCO$_{6.67}$ (p=0.123).  The original points measured by Fauqu\'e {\it et  al}\cite{Fauque} have been averaged out.  A sloping background, representing a possible linear drift of $R^{-1}_{0}$  (dashed red line), has been added to Fauqu\'e {\it et  al}\cite{Fauque} data. 
}
\label{011}
\end{figure}

$R^{-1}_{\rm meas}$ contains two terms. The first one represents the ratio between the expected magnetic intensity ${I_{mag}}$=$|F_M|^2$  and the nuclear intensity of a given Bragg peak, $|F_N|^2$. This term is intrinsic and should be comparable among the different studies, depending solely on doping and temperature as the pseudogap phase\cite{CC-review,CC-review2}.  That ratio is found, in detwinned samples, to be $\simeq $ 0.1\% on $\bf Q$=(100) (see {\it e.g.} Fig. \ref{croft-lucile}.b) \cite{Lucile17} and  $\simeq $ 0.2\% for $\bf Q$=(011)  (see {\it e.g.} Fig. \ref{011})  \cite{Fauque,comment}.  By comparing the amplitude of this term, we pointed out several drawbacks in  Croft {\it et al} analysis and calibration \cite{comment}. 

We here focus on the second term in Eq. \ref{Imag}, the so-called the bare inverse flipping ratio $R^{-1}_{0}$ \cite{Baledent-YBCO,Li2011,Lucile14,CC-review}. It  is extrinsic as it depends on the experimental setup, that of course includes the instrument parameters, as well as the sample characteristics (size, mosaic, geometry, etc ...). In principle, one expects $R^{-1}_{0}$ to not vary appreciably with the temperature. That is the assumption we made in our original reports \cite{Fauque,Li2008}. However, this crude assumption is not experimentally verified at nuclear Bragg position where no magnetic signal is expected, for instance at  large momentum Bragg peaks, like $\bf Q$=(2,0,L) or $\bf Q$=(0,2,L)  (L=0,1)  where the usual fast decay of the magnetic form factor washes out any magnetic scattering and where the IUC magnetism has never  been reported in YBCO \cite{Mook,Baledent-YBCO,Lucile15,Lucile17}.  One observes a slight change of $R^{-1}_{0} (T)$ with temperature which is linear at the first order (Fig. \ref{croft-lucile}.a). First, it should be stressed that taking a constant  $R^{-1}_{0}$ in our original reports \cite{Fauque,Li2008} does not affect our conclusion qualitatively as these data show large enough statistics.  Second, Croft {\it et al} \cite{Hayden,reply} deny this linear drift which has a huge impact on  their conclusion due to the limited statistics of their data. 

To further  assess this issue, we then plot in Fig. \ref{croft-lucile}.a the temperature dependence of $R^{-1}_{0} (T)$ obtained in various experiments.  We compared results of Croft {\it et al}\cite{Hayden} using two detwinned YBCO samples with a mass of 18 mg with our data \cite{Lucile17} where the sample was made of an array of 180 co-aligned detwinned YBCO single crystals with a mass of 3000 mg \cite{Hinkov}. 

$R^{-1}_{0} (T)$ is shown of Fig. \ref{croft-lucile}.a for our detwinned YBCO samples at Bragg positions Q=(2,0,0) or Q=(0,2,0)  (open red symbols) showing a slight linear drift.  On the same graph, one also reports  $R^{-1}_{0} (T)$  measured  for Q=(0,2,0) by  Croft {\it et al}\cite{Hayden} on their two samples. That corresponds to their best result, quoted ``With re-aligment'' in \cite{reply}. Clearly, Fig. \ref{croft-lucile}.a  shows the superior data quality of our data \cite{Lucile17} in all respects -  better polarization, six times better error bars and less sloping background - indicating a better reproducibility  and stability of our data.  Typically,  the average slope of $R^{-1}_{0} (T)$ in our data  is always more than twice lower than in the data of Croft {\it et al}  (Fig. \ref{croft-lucile}.a). Their sample realignment method gives a larger drift than our measurement with no lattice parameter re-alignment. That leads to two important conclusions: i) clearly, the argument that they remove the drift by a sample realignment method is erroneous, ii) the fact we do not re-align the lattice parameter in our measurement \cite{CC-review} is not causing the ``inevitable drift" (as we state in \cite{comment}). Of course, Croft {\it et al}\cite{reply} data 'with ``No re-aligment'' (dashed green line in Fig. \ref{croft-lucile}.a)  exhibits an even worst situation with 10 times larger variations totally incompatible with our data.  Fig. \ref{croft-lucile}.a  demonstrates our much better data reproducibility, an essential asset to detect the IUC magnetism.

As a result, the suggestion of Croft {\it et al}\cite{reply} that our  “continuous temperature run” method would produce artifacts with linear variations of $R^{-1}_{0} (T)$ due to changes in the lattice parameter is clearly disproved by Fig. \ref{croft-lucile}.a, otherwise our data should show much larger slope than the best (''with re-alignement'') ones of Croft {\it et al}\cite{Hayden,reply}. 
Similarly, sample rocking scans re-alignement cannot be the cause of the linear drift contrary to the claims in \cite{Hayden,reply}. Actually, it should be stressed that such rocking scans were also performed in our measurements when it was needed\cite{comment}, namely when the Bragg peak intensity in the NSF channel was departing from the expected Debye-Waller factor. All our measurements were otherwise performed with minimal instrument movement in order to maximize the stability of the experimental conditions contrary to \cite{Hayden,reply}. That may explain the better reproducibility of our data. Finally, although it cannot be due to  the lattice parameter or mosaic change with temperature, the weak temperature drift of $R^{-1}_{0} (T)$ is present in all data including those reported by  Croft {\it et al} \cite{Hayden,reply}.

It is worth recalling that the IUC magnetism is observed at certain Bragg positions such as $\bf Q$=(10L) or $\bf Q$=(01L) and absent otherwise for $\bf Q$=(20L) or $\bf Q$=(02L) (L=0,1,2) as it is shown for L=0 in  Fig. \ref{croft-lucile}.b or at $\bf Q$=(00L)\cite{Fauque}. That feature, alone,  proves that the observed signal cannot be parasitic due to sample misalignment, for instance Bragg peaks at larger $\bf Q$, like $\bf Q$=(200)  would be more sensitive to lattice parameter changes than Bragg peak at lower $\bf  Q$, like  $\bf Q$=(100).  For instance, using the measured $a$  lattice parameter thermal change in YBCO\cite{bozin} and our incident neutron energy of E$_I$=13.7 meV, one expects a change in the Bragg scattering angle 2$\theta$  of 0.14$^\circ$ and only 0.06$^\circ$  respectively for both Bragg positions when cooling from room temperature to 100 K. That means more than twice larger shifts in 2$\theta$ scans for $\bf Q$=(200) than at $\bf Q$=(100)\cite{comment}. In case this effect could be important, one would  then expect a larger slope at $\bf Q$=(200). This is clearly not observed  (see e.g. Fig. \ref{croft-lucile}.b).  Therefore, the fact that one experimentally observes  (Fig. \ref{croft-lucile}.b) a  signal  at $\bf Q$=(100)  (due to the IUC magnetism) but not at $\bf Q$=(200)  rules out the sample misalignment parasitic effect.

In order to observe the IUC magnetism amplitude, the major experimental difficulty or complications\cite{reply} is to minimize the temperature dependence of $R^{-1}_{0}(T)$, namely $\delta_T R^{-1}_{0}(T)$. That should be noticeably lower than the ratio of the amplitude of the expected magnetic signal to the nuclear intensity in order to get relevant data, yielding $\delta_T R^{-1}_{0}(T) << $ 0.1-0.2 \%.  As shown by  Fig. \ref{croft-lucile}, the thermal drift between 300 K and 100 K  is typically $\simeq $ 0.01\% from our data and $\simeq$ 0.05\% for the data of Croft {\it et al}\cite{Hayden}.  That systematic uncertainty adds up with  the statistical data uncertainty of  
$ R^{-1}_{0}$, which is also $\simeq \pm$ 0.05\% for the data of Croft {\it et al}\cite{Hayden}. Comparing these numbers,  it  is obvious that the insufficient data reproducibility of Croft {\it et al}\cite{Hayden} prevents from observing the IUC magnetism\cite{comment}.   In contrast, Fig. \ref{croft-lucile}.b also shows that our inevitable extrinsic drift  $\delta_T R^{-1}_{0} (T)$ is marginal compared to the  IUC magnetism amplitude.  Our data  have enough accuracy to highlight the IUC magnetism contrary to those reported by Croft {\it et al}\cite{Hayden,reply}.

\section{Polarization homogeneity and sample vertical displacement}

The value of $R^{-1}_{0}$ and its thermal drift is necessarilly extrinsic as  it depends on the experimental setup.
 We here show that this unwanted drift  of $R^{-1}_{0} (T)$  can be due to the vertical displacement of the sample when changing the temperature within an inhomogeneous neutron spin polarized beam, a point we already stressed  previously \cite{Baledent-YBCO,Li2011,Yang2018,Lucile14,CC-review,comment}.

The inhomogeneities in the neutron beam polarization are related to the combination of imperfections of all the devices used to polarize, maintain (guide field) and analyse the neutron spin polarization. On the instrument IN20 at Institut Laue Langevin (ILL, 54 MW reactor), used in \cite{Hayden}, Heusler crystals are  used to polarize and analyse the neutron spin polarization. On the intruments 4F1 at LLB (Orph\'ee, 14 MW reactor) that we extensively used  \cite{Fauque,Mook,Baledent-YBCO,Lucile17,Li2008,Li2011,Baledent-LSCO,Lucile14,CC-review} and D7 at ILL\cite{Lucile15,Yang2018}, the neutrons are polarized by a multilayer supermirror (bender) made of CoFeV/TiZr coatings with  m=2.5 on 4F1 \cite{gatchina} and Co/Ti layers  m=2.8 on D7 \cite{Stewart}. The polarization is next analyzed on 4F1 by Heusler crystals whereas a larger bender bank is installed on D7\cite{Stewart}. Therefore, each instrument would exhibit different polarization inhomogeneities as Heusler crystals and bender devices are characterized by different parameters. As the Heusler monochromator and analyzer consist of large crystals array  of co-aligned crystals, the polarization given by diffraction depends on the local orientation and mosaic spread inhomogeneities and can lead to variations of up to a few percents\cite{Hayden}.  In contrast, supermirrors are made of alternating layers of materials that  provide efficient reflection neutrons at angles exceeding the critical angles of the materials. Typically, the polarizing efficiency and transmittance of such a bender (section horizontal x vertical=25x80 mm$^2$) are not noticeably different at positions taken over the height of the entrance window \cite{gatchina2}.  Further, the polarization homogeneity would depend on the neutron guide field system in general and in particular on the Helmholtz-like coils used to control the neutron polarization direction at the sample position by applying a small 10-15 Gauss field. Therefore, the neutron beam polarization inhomogeneities would depend on the different intrument configuration and can vary substantially between the different instruments used. 

Being one element present in the neutron path, that enters into $R^{-1}_{0} (T)$, the sample geometry and mosaicity also contribute to the final polarization.  The polarization is typically larger for small sample than for larger one because of the spatial distribution of polarization.  It is actually remarkable that typically larger flipping ratio ($\sim$ 40) can be readily achieved on 4F1 with much larger sample ($\sim$ cubic centimeter) than on IN20 with smaller samples (2x3x0.5 mm$^3$). That proves a spatially more homegeneous polarization distribution on 4F1 over a larger beam section.

We agree with Croft {\it et al}\cite{reply} that inhomogeneities in neutron beam polarization are inevitable, a point we have previously and continuously re-assessed  \cite{Baledent-YBCO,Li2011,Yang2018,Lucile14,CC-review,comment}. From the previous section, this leads to a systematic extrinsic thermal drift of the inverse flipping ratio, although its amplitude can noticeably vary among experiments.  One relevant question is what can be the cause for it ?   

Croft {\it et al}\cite{reply} wrongly attribute it to the change of lattice parameter and sample orientation with temperature. That scenario does not hold (see above) simply because that effect is smaller than the broad momentum resolution of the instrument. As the sample size and mosaic enter into the instrument resolution, such an effect would be noticeably reduced by using larger samples. With a sample of only 2 mm size in the horizontal plane, Croft {\it et al}\cite{reply}  face a much more difficult situation as the neutron spot on the analyzer is much smaller than the analyzer size (horizontal × vertical: 200x100 mm$^2$ on IN20\cite{Hayden}). It then experiences variable  polarization of the Heusler crystals. On 4F1, we use a smaller Heusler analyzer  for diffraction studies (horizontal × vertical: 100x40 mm$^2$) and larger samples (10-15 mm size); the polarization of the beam spot (typically 20x20 mm$^2$) is then better averaged on the  Heusler crystals of the analyzer. Further,  in contrast with \cite{Hayden,reply}, we use samples with broad mosaicities ($\sim$ 1 deg)  which also average the beam polarization.  Accordingly, this implies that the polarization variation when rotating sample or moving the Bragg angle 2$\theta$ (as shown by Fig. 6 in \cite{Hayden})  is less important in our measurement.  As a final result, the thermal drift  of $R^{-1}_{0}$  is also less significant  in our measurement as clearly demonstrated in Fig. \ref{croft-lucile}.a.

In these experiments covering a large temperature range (from 50 K to 300 K), the sample is typically attached to the cold head of closed-cycle cryogenerator at LLB and at the end of the  sample stick of a cryostat at ILL. Due to thermal dilatation of aluminium present in the cold region, the sample moves vertically upon cooling. This effect is obviously less important if the cold region around the sample is limited: for instance, one would expect a more prominent effect for a cryostat than for a closed-cycle  cryogenerator. 

The Al thermal expansion is known\cite{Al-thermal} to be $\delta l/l$= 40x10$^{-4}$ between 100 K and 300 K. Assuming 25 cm of the Al sample stick in the cold region,  one obtains a sample displacement of about 1 mm when cooling from 300 K to 100 K. This is not negligible compared to the sample size (especially in \cite{Hayden} where the sample vertical size  was only 3 mm). That effect was not compensated in the aligment protocol  celebrated in \cite{Hayden}. Therefore, the sample would experience a varying polarization upon cooling as the neutron beam polarization is not spatially homogenous. Clearly, that could contribute to the observed drift of the neutron polarization,  which corresponds to a variation of about $\delta R^{-1}_{0} \sim 0.005 - 0.05 \%$ between 100 K to 300 K for the various experiments shown in Fig. \ref{croft-lucile}.a.  In case that this variation is due to the sample vertical displacement, that would correspond to a vertical spatial gradient of $R^{-1}_{0}$ of $\sim 0.005 - 0.05 \%$/mm, a  realistic value. Therefore, a major cause for the thermal drift  of $R^{-1}_{0}$ could be the vertical sample displacement. In this scenario, it should be present in any PND experiments. Further,  it would be more detrimental  when using smaller samples. In details, it would also depend on the device used to cool the sample (cryostat or cryogenerator).  The reported experiments  (Fig. \ref{croft-lucile}.a) are consistent with the trend expected from all these features. 

\section{Consistency between all PND experiments}

Croft {\it et al}\cite{Hayden,reply} claimed that their data show no evidence of the IUC magnetism. This negative statement is questionable as their data statistics, reproducibility (see Fig.  \ref{croft-lucile}) and analysis show severe limitations: actually only one result, for the Q=(011) reflection, shows  an apparent disagreement between both datasets \cite{comment}.  For the Bragg  Q=(010)  position, we have already demonstrated \cite{comment} that both measured data fully agree (a point not contested in \cite{reply}) even if Croft {\it et al}\cite{Hayden} data did not have sufficient statistics to observe the IUC magnetism.

The magnetic IUC intensity, $I_{mag}$, can be determined using Eq. \ref{Imag} only once the background baseline is known with enough confidence. Fig. \ref{croft-lucile}.a  demonstrates that we had a much better control of 
the thermal drift of  $R^{-1}_{0}(T)$. Croft {\it et al}\cite{Hayden,reply} denied this thermal drift by arbitrarily fitting  $R^{-1}_{0}(T)$ by a constant. This choice is at odd with the notion of polarization inhomogeneities in the neutron beam on which there is a consensus. Their point of view is inconsistent with the inevitable vertical shift of the sample with temperature. Our criticisms on ref. \cite{Hayden} are thus founded. It has important consequences \cite{comment} that  the final results in Fig. 11 in \cite{Hayden} and Fig. 3  in \cite{reply} are not correct as the reported data points  have very large systematic errors (in addition to the statistics) related to the arbitrary choice of  $R^{-1}_{0}(T)$. We have considered consistent scenarios  showing that Croft {\it et al}\cite{Hayden} data accuracy was insufficient to be able to detect the IUC magnetic signal \cite{comment}. In particular, we have proposed that if the IUC magnetism is characterized by a finite correlation length, it will  further  reduce the magnetic signal in their experiment  by a factor 2 to 3 depending on the sample mosaicity\cite{note}. In their reply, Croft {\it et al}\cite{reply} rejected these scenarios based only on their inconsistent assumption of a flat thermal variation of $R^{-1}_{0}(T)$. 

In Fig. \ref{011}, we  show that a consistent linear variation of $R^{-1}_{0}(T)$ gives an agreement between both datasets for the $\bf Q$=(011) reflection.  We take a sloping variation of $R^{-1}_{0}(T)$ shown  by the two points measured at high temperature (dashed red line in Fig. \ref{011}). Due to the scarce number of points measured by Croft {\it et al}\cite{Hayden},  this choice is  as arbitrary as the  unrealistic choice of a constant  $R^{-1}_{0}(T)$ they made.   In  Fig. \ref{011}, we next add this linear background to the measured points of Fauqu\'e {\it et  al}\cite{Fauque} and compare with the raw data for the samples measured by Croft {\it et al}\cite{Hayden}. Both datasets fully agree with no further ajustable parameters as the intrinsic IUC amplitude is known \cite{Fauque}. We point out that this analysis is as relevant as the negative conclusion pushed forward by Croft {\it et al}\cite{Hayden}.

Therefore,  none of the data reported by  Croft {\it et al} challenges the universality of the intra-unit cell  magnetism in cuprates. 
Further, we recently reported a similar magnetic signal in the antiferromagnetic iridates system Sr$_2$(Ir,Rh)O$_4$\cite{jj} below a temperature different from {\rm $T_N$} suggesting that loop-current-type electronic states can exist in other  oxides.

{\bf acknowlegments}

We acknowledge financial supports from the project NirvAna (contract ANR-14-OHRI-0010) of the Agence Nationale de la Recherche (ANR) French agency.


\begin{thebibliography}{9}
\bibitem{note} The definition of the scattering function in \cite{reply} is inconsistent with the correlation length we have been using in \cite{Mook}. 
 As a result, the quoted relative correction 1.4 in \cite{reply} (which applies to the correction of the data when the IUC magnetism has a finite correlation length) is not correct.
\end{thebibliography}
\end{document}